# On the nature of antiferromagnetism in the $CuO_2$ planes of oxide superconductors

J. Friedel* and M. Kohmoto**

* Physique des Solides, CNRS and Université Paris-Sud – 91405 Orsay (France)

** Institute for Solid State Physics, University of Tokyo (Japan)

**Summary**

Recent results on electrons and holes doped $CuO_2$ planes confirm the marked covalency of CuO bonding, suggesting a band picture of long and short range antiferromagnetism. The maxima of superconductive $T_c$ versus doping can be related to the crossing by the Fermi level of the edges of the pseudogap due to antiferromagnetic short range order (bonding edge for holes doping, antibonding one for electrons doping). The symmetry of the superconductive gap can be related to the Bragg scattering of electronic Bloch states near the edges of the AF pseudogap. Assuming a standard phonon coupling, one then predicts for commensurate AF a pure d symmetry of the superconductive gap for underdoped samples and d symmetry plus an ip contribution increasing linearly with overdoping. This seems in agreement with recent measurements of gap symmetry for YBCO, but should be more fully tested, especially for electron doped samples. The simple band approximation used here could no doubt be made more realistic by a specific inclusion of electron correlations and by a better description of AF short range order. Uncommensurate AF, as in LSCO, is not considered here.

———

The purpose of this paper is to propose an approximate but original description of superconductivity in the cuprates. It uses a number of concepts which were successively considered by one of the authors in this field and which will be first recalled [1] to [4]. These concern the covalency in the $CuO_2$ planes and the pseudogap due to antiferromagnetic short range order. We feel it is the simplest if not the only way to explain two recent sets of experiments, which have stimulated anew our interest in the field : they confirm the qualitative symmetry of the phase diagrams of electrons and holes doped compounds [5], [6], [7] and the symmetry of he superconductive gap as a function of doping [8]. The essential conclusion is that the maxima of the critical superconductive temperature occur when the Fermi level crosses one of the peaks of density of states associated with the AF pseudo gap. Stated in this way, this suggestion was made privately some years ago by the author to D. Pines, but never published ; and it will be explained why. Finally, for many reasons, this is but a crude caricature that could however be worked on to develop the model and to include various specific characteristics of some of the cuprates.

*a) Three types of LCAO pictures* [1] to [4].

In the simplest linear combination of atomic orbitals (LCAO) picture for the $CuO_2$ planes common to all cuprates, three parameters are involved : the transfer integral t between neighbouring Cu3d and O2p atomic functions, their energy difference $\Delta = E_{3d} - E_{2p}$ and the intraatomic repulsion U between electrons on a Cu ion. This simple approach neglects small possible distortions from a square lattice. It also neglects interactions of a $CuO_2$ plane with its surrounding, especially the coupling between parallel planes responsible for three dimensional (static) superconductivity : this essential part [2] will have to be introduced as a correction.

Three situations have been envisaged within these limits.

**1.- Ionicity** :    $\gg U \gg t$.

This starting point in many early discussions does not fit with the presence of holes in O2p shells, increasing in number with holes doping, as shown by X-rays and NMR techniques [9].

**2.- Charge transfer** : $U \gg$    $\gg t$.

This fits better with the doping of holes in O2p shells, and agrees with the electronic structure of small clusters [10]. However it does not fit with the observation of holes in O2p shells even in non doped samples [9]. It is also difficult to explain in this way the qualitative but striking symmetry observed in the phase diagrams for electron and hole doping of the $CuO_2$ planes (fig. 1). It would indeed be difficult to explain the symmetry observed in antiferromagnetism (AF) and supraconductivity (S) [5], [6], as well as the change from holes to electrons Hall conductivity beyond the optimum dopings $(z_o, z_o)$ for maximum superconductive Tc [6], [7]. For the basic electronic structures are very different in this charge transfer model : less than one hole per Cu ion in the electrons doped samples ; but, in the holes doped samples, one hole per Cu ion and the excess holes distributed on the O ions. The observed qualitative symmetry is especially clear in recent experiments where carriers are introduced by an electric voltage [8], thus avoiding possible distortions of phase diagram due to differences in chemical dopings for holes and electrons. Finally, if $U \gg$  , the presence of one hole on each Cu ion in holes doped samples would block effectively the motion of O2p holes if other types of transfer, between O2p orbitals, were not invoked [11].

**3. Covalency** : $U \gtrsim w = t$      with   $\gg 1$.

w is here the effective width of the covalent band built with the Cu 3d and O2p orbitals. This condition is actually fullfilled by the studies on clusters mentionned above [10], owing to the fact that each Cu has more neighbours in a $CuO_2$ plane than in the clusters considered by [3].

In this case, O2p holes should be present in sizeable amounts for zero doping and also for reasonable amounts of electrons doping. And the qualitative symmetry of phase diagrams for electrons and holes doped compounds results naturally from the near symmetry of the Cu3d-O2p bonding developed in this limit.

We shall use this picture as the most reasonable at present. It fits with the fact that, in Pauling's electro-negativity scale, copper, at the end of the 3d transitional series, is not very far from oxygen. This scheme would obviously be usefully confirmed if O2p holes were also observed in electron doped compounds (e.g; by X rays absorption spectra) ; or if the qualitative symmetry observed for magnetism, superconductivity and antisymmetry for Hall transport was extended as we expect to other properties discussed below : AF short range order, pseudogap, changes in symmetry with doping of the superconductive gap.

*b) Ionocovalency. Correlation effects* [2] [3].

First, in the <u>uncorrelated</u> picture where U is neglected, the general tight binding equations give, for the partially filled antibonding band :

$$E_k = E_d + \frac{1}{2}\left[\Delta + \sqrt{\Delta^2 + 16t^2\left(\sin^2\frac{k_x a}{2} + \sin^2\frac{k_y a}{2}\right)}\right]^{1/2} \quad (1)$$

with a width :

$$w = \frac{1}{2}\left[(\Delta^2 + 32t^2)^{1/2} - |\Delta|\right] \quad (2)$$

The Fermi level for zero doping lies along the square

$$\cos k_x a + \cos k_y a = 0 \quad (3)$$

For a pure covalent limit $\Delta = 0$

$$E_k = E_d + 2t\left[\sin^2\frac{k_x a}{2} + \sin^2\frac{k_y a}{2}\right]^{1/2} \quad (4)$$

and

$$w = 2\sqrt{2}\, t \quad (5)$$

The Cu 3d and O2p states play equivalent roles, so that the holes are equally distributed between them. For zero doping, each Cu ion has 0.5 holes and each O ion 0.25. This is to be compared with the ionic limit $\Delta \gg w$, where :

$$E_k \simeq E_d + \Delta + \frac{4t^2}{\Delta}\left[\sin^2\frac{k_x a}{2} + \sin^2\frac{k_y a}{2}\right]^{1/2}$$

with a (smaller) width :

$$w \simeq \frac{4\sqrt{2}\, t^2}{\Delta}$$

Computations on aggregates suggest an approximate covalent limit, where $\Delta$ is positive and of order $2t$, thus, according to (2), of the same order as $w$ [10] [3]. Equation (1) then shows that the equirepartition of holes between Cu and O is still approximately preserved near the middle of the antibonding band, thus practically near the Fermi level. Near the edges of that band, pure 2p character prevails at the bottom and pure 3d at the top, over a width $ka \simeq \Delta/2t$. As the area $4\pi^2/a^2$ corresponds to one hole per cell in undoped compounds, the pocket of 2p character at the bottom of the band contains $\simeq k^2 a^2/4\pi^2 \simeq \Delta^2/16\pi^2 t^2$ Cu 3d holes per cell. For $\Delta \simeq 2t$, this is a small fraction of unity. There shall therefore be a small and constant shift of the holes from O to Cu, independent of doping when $\Delta$ increases from 0 to its likely value of order $2t$.

We should now include <u>electrons correlations</u> by considering $U \neq 0$, but shall at the moment neglect the (important) magnetic consequences, to be discussed later. The effect of $U$ will be to reduce the charge fluctuations on the Cu ions. This will decrease somewhat the effective width of the band considered, without however changing the delocalised electrons picture. In fact, these electrons correlations should have small consequences. The same applies to the d band of transitional metals such as Ni, Pd, Pt, with also a fraction of d holes per atom : in the limit of small number of d holes per atom, it is known that the Coulomb repulsion cannot, as such, create a Mott-Verwey insulator ; and the correlations only reduce the large charge fluctuations without changing much the cohesive properties. Indeed, the correlations can be described in terms of an effective S matrix [12] :

$$U_e \simeq U/(1 + U/w_o)$$

always smaller than the band width $w_o$ of the holes ; and cohesion and magnetic properties can be satisfactorily described in small power developments in $U_e$ [13].

*c) Long range antiferromagnetism* [2]

We consider now the long range antiferromagnetism observed for small dopings of holes or electrons (fig. 1).

For <u>zero doping</u>, we shall assume with Lomer that the effect of $U_e$ is to stabilise an antiferromagnetism of wave vector $\underline{Q}_o$, the size of the square of the Fermi surface (fig. 2). This antiferromagnetism commensurate with the lattice will induce a gap in the density of states, producing a band insulator. The perfect nesting of the Fermi sheets by the $\underline{Q}_o$ translation corresponds to a strong instability, thus a priori rather large moments on the copper ions. The total atomic moments developped cannot be larger than the average number of holes on copper ions, which is itself not much larger than 0.5. Observed atomic moments $\mu$ of order 0.5 $\mu_B$ are therefore not surprising. In the approximation of our model and within an extended Hartree-Fock scheme, such moments lead to corrections $\pm U_e\mu/\mu_B$ to the copper atomic potential, depending on the relative directions of the electronic spin and the atomic moment considered [14]. Indeed the observed Néel temperature $T_N$ and the AF gap are compatible with $U_e \lesssim w$ and a perturbation treatment of $U_e$. The situation above $T_N$ probably involves AF fluctuations and possibly an Anderson localisation by magnetic disorder [2].

For <u>finite doping</u>, Lomer's argument would lead to a static incommensurate antiferromagnetism, with a wave vector $\underline{Q}$ varying essentially linearly with doping, so as to follow the size of the Fermi surface. However in most cases, the antiferromagnetism keeps <u>commensurate</u> with the lattice, with the wave vector $\underline{Q}_o$. This will be the only case discussed here. It can be understood as due to the very sharp and quasionedimensional feature of the peak in density of states at the edges of the AF gap, near the square of size $\underline{Q}_o$ (fig. 2 and 3). For this commensurate AF, increasing numbers of carriers due to doping should shift the Fermi surface away from the AF gap. However, because of the large gap and its very sharp peaks of density of states, one can expect in this range the carriers to be preferentially captured by doping imperfections and not to carry current. In the recent experiments of doping by electrical voltage, one could similarly think injected carrier to be captured by imperfections or the applied voltage to be smaller than the AF gap : the doped AF phase should again be insulating.

More precisely, in the presence of an antiferromagnetism with a magnetic potential alternating from Cu to Cu ion, the energy of the band electrons becomes :

$$E_{\underline{K}} = \frac{1}{2}(E_{\underline{k}} + E_{\underline{k}-\underline{Q}_o}) \pm \frac{1}{2}[(E_{\underline{k}} - E_{\underline{k}-\underline{Q}_o})^2 + 4\ v\ ^2]^{1/2} \qquad (6)$$

where $v = \langle \underline{k}\ |v|\ \underline{k} - \underline{Q}_o\rangle$ is the matrix element of the atomic potential $\pm U_e\mu/\mu_B$ due to antiferromagnetism. The corresponding wave function is :

$$|\tilde{K}\rangle = \sum_k \alpha_k |\tilde{k}\rangle + \sum_k \beta_k |\tilde{k} - \tilde{Q}_o\rangle \qquad (7)$$

with :

$$\frac{\alpha_{\tilde{k}}}{\beta_{\tilde{k}}} = -\frac{E_{\tilde{K}} - E_{\tilde{k}}}{\langle \tilde{k}|v|\tilde{k} - \tilde{Q}_o\rangle} \qquad (8)$$

Using the remarks above that only states near to the Fermi level are involved in the magnetic perturbation and that these states are not much sensitive to (i.e. $U_e$ and small), we can compute $E_{\tilde{K}}$ in the extreme covalent limit $= 0$ :

$$E_{\tilde{K}} \simeq E_d + 2\sqrt{2}\ t \pm [t^2/(\cos k_x a + \cos k_y a)^2 + {}^2]^{1/2} \qquad (9)$$

The AF gap, of width $2\ v$ , is centred at energy

$$E_o = E_d + 2\sqrt{2}\ t$$

as pictured fig. 3. Development along the side AB, fig. 2, gives with :

$$k_x a + k_y a = -\ ,\qquad 0$$

$$k_x a - k_y a = u \qquad (10)$$

$$E_{\tilde{K}} \simeq E_o \pm\ v\ \pm \frac{t^2}{|v|}\ 2\cos^2 \frac{u}{2} \qquad (11)$$

The corresponding density of states diverges at the gap edges. Thus, if $E$ is the distance of $E$ to the gap,

$$n(E) \simeq \text{const}\ \left(\frac{|v|}{t^2| E|}\right)^{1/2}\ \ell n\ \frac{\text{const}\ t^2}{|v|\ E|} \qquad (12)$$

The dominating contribution comes from the Van Hove anomaly at the corners A, B of the square, fig. 2.

Thus, near A :

$$E_{\tilde{K}} \simeq E_o \pm\ v\ \pm \frac{t^2}{4|v|}\ \left(k_x^2 a^2 - (\ -k_y a)^2\right)^2 \qquad (13)$$

This van Hove anomaly is much flatter in energy than that at the corners A, B for non magnetic phase, when the density of states only diverges logarithmically in $E$ [26]. The divergence in $E$ is also somewhat larger than the contributions to n(E) along the side AB of the square, where, as in a one dimensional problem [27], they diverge as $E^{-1/2}$. This striking result comes from the square geometry of the Fermi surface of non doped samples ; it is also found in the ionic limit, if one assumes also electrons to be delocalised. Finally, on either side of the AF gap, the surfaces of constant energy have the same form in reciprocal space as with no gap (fig. 2) but repeated in all the squares neighbouring ABCD and with a reduced energy scale.

The stability of the AF gap can be approximately measured by summing up the one electron energies of the occupied states :

$$\Delta E = \int^{E_F} n(E)\, E\, dE - \int^{E_{F_o}} n_o(E)\, E\, dE \qquad (14)$$

where $E_F$ and $E_{F_o}$ are the Fermi levels in presence and in absence of antiferromagnetism respectively ; $n(E)$ and $n_o(E)$ are the corresponding densities of states counted per $CuO_2$ in a plane [4]. In (14), a self energy term in $U_e$ is counted twice and should be subtracted. This affects the absolute stability of the AF phase, but not the relative stability of magnetic phases with similar magnetic moments.

The average number of doping electrons per $CuO_2$ is :

$$z = \int^{E_F} n(E)\, dE - \int^{E_{F_o}} n_o(E)\, dE \qquad (15)$$

This gives :

$$\frac{d\,\Delta E}{dz} = E_F - E_{F_o} \qquad (16)$$

$$\frac{d^2\,\Delta E}{dz^2} = \frac{1}{n(E_F)} - \frac{1}{n_o(E_{F_o})}$$

It is then easy to check that, for a commensurate AF $\tilde{Q}_o$, $\Delta E(\tilde{Q}_o, z)$ has a minimum at zero doping, with a break in slope $d\Delta E/dz$ (related to the gap $2\,\Delta v$, thus to $U_e\mu/\mu_B$) and a negative curvature (due to the peak in $n(E)$ near the gap edges). The near symmetry of $\Delta E(\tilde{Q}_o, z)$ for electron and hole doping is related to that of $n(E)$, fig. 3. This is related to the quasi one dimension of $E_{\tilde{k}}$ near zero doping and to the slow variation with doping of the magnetic moment $\mu$.

For an <u>uncommensurate</u> AF at $\tilde{Q} \neq \tilde{Q}_o$, the gap of $n(E)$ is expected to be smaller, thus also the stability of the corresponding AF phase, owing to the increasing two dimensional nature of $E_{\tilde{k}}$. This gives obviously a stability $\Delta E(\tilde{Q}_o, z)$ less marked and more asymmetrical, which can be expected to be usually above $\Delta E(\tilde{Q}_R, z)$ in the region of stability of the AF order (fig. 4).

d) <u>Short range antiferromagnetic order</u>. [1] [2] [4]

Ever since a strong antiferromagnetic short range order was observed by neutron scattering and Mössbauer effect in a large range of <u>hole doped</u> oxide superconductors such as YBCO, it has been clear that this phenomenon should play an important role in the phase diagram of these materials. In particular, up to the optimum doping $z_o$ and beyond, this short range is strong at temperatures well above the maximum of Tc and its possible effect on superconductivity must be taken into account. We shall restrict ourselves here to compounds with <u>commensurate</u> short range AF, thus will not consider the somewhat exceptionnal case of LSCO.

The first idea along these lines has been to consider a coupling scheeme of the electrons via AF fluctuations [15]. And the observation of a superconductive gap with d symmetry was apparently in agreement with this picture [16].

We shall take a different point of view, stressing first the special properties and structures associated with this short range AF. Discussion of the effect on the superconductive gap will be deleted to the last paragraph.

The main idea, first proposed in this field by one of the authors [1] to [3] is that, when a long range AF is replaced by a short range one of similar strength, the gap g in the density of states should be replaced by a <u>pseudogap</u> g , as pictured fig. 3 : the disorder should round off the peaks B and AB of density of states ; and, if disorder is sufficient, a continuous trail of states, with a density decreasing towards the middle of the gap should appear. Such states in the pseudogap should be primarily made of parts of evanescent Bloch functions of the real gap connected together in the regions of disorder, in such a way that, except possibly for Andersen localisation near the middle of the pseudogap, such states should be conducting. Such a qualitative picture of a pseudogap and indeed its name were first evolved by Mott in the early fifties, in connection with atomic short range order.

A rough image of this pseudogap can be obtained by considering a localised AF (commensurate) spin wave. Looking for simplicity's sake at the one dimensional problem, it is easy to compute the phase shifts produced on a non magnetic matrix by the presence of a length of the commensurate AF phase. These phase shifts give a good idea of the variation of density of states produced locally along the AF phase, if this is much longer than the Fermi length of the matrix, i.e. more than several interatomic units. The coherence length of the short range AF order fulfills this condition ; and one finds indeed a pseudogap as sketched figure 3. A similar result would be obtained for onedimensional antiphase domains of random sizes.

The decrease of Knight shift, thus of density of states at the Fermi level, when underdoping increases below $z_o$ [17], was first attributed to the progressive opening of a pseudogap due to AF fluctuations of increasing strength for decreasing doping z below $z_o$ [1]. As expected in such a picture, a peak in density of states was observed later at an excitation energy increasing from zero with z decreasing from $z_o$ ; and a corresponding anomaly was observed at an increasing temperature $T_g$, fig. 5 [18] : this excitation energy could be considered as measuring the energy distance between the bonding peak B, fig. 3 and the Fermi level, assumed above $E_B$ for $z < z_o$.

It is however clear in compounds with commensurate short range AF that this AF short range order does not decrease to zero at $z_o$, but remains strong at low temperatures well beyond this limit. The crossing of the binding peak B by the Fermi level at $z = z_o$ is therefore not so much due to a closing of the pseudogap with increasing doping as to <u>a shift of the Fermi level associated with increasing hole doping</u>. It is this (new) point of view that we wish to develop in this paper.

If we neglect the variation with doping of the amplitude of magnetic moments, an analysis of stability based on equations (13) to (16) for a density of states as sketched fig. 3, leads to a stability of the pseudogap  $E(Q_o,z)$ as pictured qualitatively fig. 4. The pseudogap should be less stable than the gap for small dopings ; but, owing to the finite density of states in this pseudogap, this should become more stable than the gap at higher dopings, and before the Fermi level reaches the peak B of the pseudogap. There should therefore be a spontaneous transition from long range to short range order beyond a certain doping, through a first order transition. This model does not rely on increasing entropy by disorder ; and thus it applies at low temperatures. It explains the otherwise surprisingly large extent in doping of the short range AF. Finally it does not make any specific and detailed prediction about the fine structure of short range order.

The similarity of long range AF and superconductivity phase diagrams for electrons and holes doped compounds suggests that the same general characteristics should apply to <u>electrons doped</u> samples. This short range AF should be present beyond the long range AF (fig. 1), with atomic moments similar to those in the long range AF phase, decreasing with increasing electron doping. The characteristic features of an AF pseudogap should be observed, with a Knight shift increasing with doping up till the optimum doping $z_o$ with maximum $T_c$ ; a peak in the density of states crossing at $z_o$ the Fermi level should be observed, corresponding to the AB peak of the pseudogap.

From fig. 3, it is clear that the excitation energy from the Fermi level to the peak B of the pseudogap corresponds to hole excitations for underdoped samples $z < z_o$ and to electron excitations for the overdoped samples $z < z_o$ in the case of hole doping, while the reverse should be true for electron doped samples ($z \lesssim z_o$). This analysis suggests that, for overdoped samples, a (weaker) anomaly at a temperature $T_p$ increasing with doping from $z = z_o$ (or $z_o$) should be observed (cf. fig. 5). This is in agreement with the general picture given for hole doped samples by Varma [19].

*e) Superconductive coupling and gaps.*

Let $V(\underset{\sim}{K}, \underset{\sim}{K})$ be the effective supraconductive coupling between electrons of opposite spins, such that the superconductive gap $\Delta_{\underset{\sim}{K}}$ is given in the BCS approximation by :

$$\Delta_{\underset{\sim}{K}} = - \sum_{\underset{\sim}{K}} \frac{V_{\underset{\sim}{K}\underset{\sim}{K}} \Delta_{\underset{\sim}{K}}}{2\sqrt{\varepsilon_{\underset{\sim}{K}}^2 + \Delta_{\underset{\sim}{K}}^2}} \tanh \sqrt{\frac{\varepsilon_{\underset{\sim}{K}}^2 + \Delta_{\underset{\sim}{K}}^2}{2 k_B T}} \qquad (17)$$

As usual, $\varepsilon_{\underset{\sim}{K}}$ is the one particle energy measured from the Fermi level. Because of the Bragg scattering of the Bloch functions $|\underset{\sim}{k}\rangle$ by the short range AF order, the wave functions $|\underset{\sim}{K}\rangle$ to be used are combinations of such Bloch functions. We shall assume, in this order of magnitude estimate, that they are the same combinations (7) (8) as in the long range order.

From (7)

$$\langle \underset{\sim}{K} | V | \underset{\sim}{K} \rangle = \alpha_{\underset{\sim}{k}} \alpha_{\underset{\sim}{k}} \langle \underset{\sim}{k} | V | \underset{\sim}{k} \rangle + \alpha_{\underset{\sim}{k}} \beta_{\underset{\sim}{k}} \langle \underset{\sim}{k} - \underset{\sim}{Q}_o | V | \underset{\sim}{k} \rangle$$

$$+ \beta_{\underset{\sim}{k}} \alpha_{\underset{\sim}{k}} \langle \underset{\sim}{k} | V | \underset{\sim}{k} - \underset{\sim}{Q}_o \rangle + \beta_{\underset{\sim}{k}} \beta_{\underset{\sim}{k}} \langle \underset{\sim}{k} - \underset{\sim}{Q}_o | V | \underset{\sim}{k} - \underset{\sim}{Q}_o \rangle \qquad (18)$$

Near the peak of the pseudogap (fig. 3), one can assume that the main contribution to $|\underset{\sim}{K}\rangle$ will be the same as at the peak of the gap in the long range AF.

For the binding peak B, this gives :

$$\alpha_{\underset{\sim}{k}} = \beta_{\underset{\sim}{k}} = 1/\sqrt{2} \qquad \text{for spin up}$$

(19)

$$\alpha_{\underset{\sim}{k}} = -\beta_{\underset{\sim}{k}} = 1/\sqrt{2} \qquad \text{for spin up}$$

if the moment developped at the origin is up.

If then the coupling V is through phonons, one can assume $\langle \underset{\sim}{k}|V|\underset{\sim}{k}'\rangle \simeq 0$ except for $\underset{\sim}{k} \simeq \underset{\sim}{k}'$, where it is negative. It is then easy to check from (18) that :

$$\langle \underset{\sim}{K}|V|\underset{\sim}{K}'\rangle \leq 0 \qquad (20)$$

$$\langle \underset{\sim}{K}|V|\underset{\sim}{K}-\underset{\sim}{Q}_o\rangle \simeq -\frac{1}{2}\langle \underset{\sim}{k}|V|\underset{\sim}{k}'\rangle \qquad (21)$$

This means that, with phonon coupling, one expects an apparent repulsive coupling as due to AF fluctuations $|\underset{\sim}{K}\rangle$ developped by Bragg scattering on the short range AF. From equation (17), one expects a gap $\Delta_{\underset{\sim}{K}}$ with d symmetry. And, as the effective density of $|\underset{\sim}{K}\rangle$ states is expected to be large along the hole side AB of the square, fig. 2, the corresponding $T_c$ is expected to be optimal, i.e. $z \simeq z_o$ (fig. 3). It is also easy to check from (18) that a coupling V through AF fluctuations, where $\langle \underset{\sim}{K}|V|\underset{\sim}{K}'\rangle \simeq 0$ except for $|\underset{\sim}{K}'\rangle = |\underset{\sim}{K} \pm \underset{\sim}{Q}_o\rangle$ (where it is positive) leads to no appreciable coupling between the dynamical function $|\underset{\sim}{K}\rangle$.

For overdoped compounds where the Fermi level shifts below the peak B of the pseudogap (fig. 3), and again using the wave functions $|\underset{\sim}{K}\rangle$ of long range AF as an approximation, the energy $E_{\underset{\sim}{k}}$ varies linearly with $\delta$ (equ. (4) and (10)) while $E_{\underset{\sim}{K}}$ varies quadratically (equ. (11). To first order in $\delta$,

$$E_{\underset{\sim}{K}} - E_{\underset{\sim}{k}} \simeq -v + \eta$$

where

$$\eta = \frac{1}{\sqrt{2}} t \cos\frac{u}{2}$$

- Thus, from (8),

$$\alpha_{\underset{\sim}{k}}/\beta_{\underset{\sim}{k}} = -\alpha_{\underset{\sim}{k}'}/\beta_{\underset{\sim}{k}'} = 1 - \eta/v$$

- Then, to second order in , equation (20) is replaced by :

$$\langle \underset{\sim}{K}|V|\underset{\sim}{K}\rangle \quad \frac{2}{|v|} \langle \underset{\sim}{k}|V|\underset{\sim}{k}\rangle \tag{22}$$

while equation (21) is unchanged.

As varies linearly with overdoping $(z - z_o)$ at the Fermi level, the gap equation (17) acquires besides its real d gap, an imaginary component, cf. [23] [25], increasing linearly with overdoping and with p symmetry due to the node of <K V K> along AB :

$$\underset{\sim}{_K} \quad \underset{\sim}{^d_K} + i \underset{\sim}{^p_K} \tag{23}$$

The strong Van Howe anomaly at the corners of the square, fig. 2 contributes little to the imaginary part because goes to zero at A and the imaginary part has a $d_{x^2-y^2}$ symmetry. In the pseudogap, both high densities of states involved in the d and p parts of the gap are reduced from the values deduced from densities of states such as (12). Also, with increasing overdoping, these densities decrease, leading to a general attenuation of $\underset{\sim}{_k}$ and of $T_c$.

Finally for <u>underdoped samples</u>, one can assume the Fermi wave functions in the pseudogap (fig. 3) to be built mostly with localised Bloch functions of complex wave vectors. Taking for instance

$$k_x a + k_y a = \quad + i$$
$$k_x - k_y = 0 \tag{24}$$

we find, using again wave functions of the long range AF,

$$E_{\underset{\sim}{K}} \quad E_d + t\sqrt{2}\ (1 + ch\frac{}{2})^{1/2} - [\ v\ ^2 - \frac{t^2 sh^2 \frac{}{2}}{(1 + ch\frac{}{2})^2}]^{1/2}$$

vanishes at the gap edge and varies linearly again with underdoping $(z_o - z)$. Thus $E_{\underset{\sim}{K}} - E_{\underset{\sim}{k}}$ only varies to second order in . To the same approximation that leads to (23), we find then a pure d superconductive gap, with equation (21) unchanged. Here, the densities of states involved in (17) decrease faster with (under)doping, leading again to a decrease of gap and $T_c$ with increasing $z_o - z$.

These predictions are in general qualitative agreement with experiments on the variation with doping of $T_c$ and with the symmetry of the superconductive gap [18]. The proposed p character in (23) seems as acceptable as the s or $d_{xy}$ envisaged so far. The fact that the p character develops progressively means, in particular, that the d and p characters have their physical origin in the same general type of coupling [20]. Finally, if our interpretation is correct, the phonon coupling $\langle \underset{\sim}{k} |V| \underset{\sim}{k} \rangle$ responsible for supraconductivity could explain the modest but real isotope effect observed in $T_c$ [2].

The same type of analysis should apply to <u>electron doped</u> compounds if, as we believe, short range AF order develops beyond the observed long range AF one. Here, it would be the AB peak of the pseudogap, fig. 3, which would be involved ; and it is easy to show that the same conclusions apply to the symmetry of the superconductive gap and to the maximum of $T_c$ for the crossing by the Fermi level of the (AB) peak of the pseudogap at $z = z_o$. As magnetic moments are expected to decrease slowly with electron doping, the pseudogap should be here less marked in width and in height of its peaks. Thus $T_c(z)$ is expected to have a maximum value $T_c(z_o)$ smaller than for hole doped compounds, as is indeed the case (fig. 1). A superconductive gap with $d_{x^2-y^2}$ symmetry has also been observed in underdoped electrons compounds [21]. But more systematic studies are obviously in order in this range, to test in particular an ip character of the gap increasing linearly with overdoping.

## Conclusions

We present an approximate description of superconductivity in cuprates which we feel has some original aspects. We stress that, if covalency is notable in the $CuO_2$ planes, as the symmetry between electrons and holes doping suggests, holes are present in very sizeable amounts in O2p as well as Cu3d orbitals, for electrons doped as well as hole doped compounds. In a situation somewhat similar to that of transitional metals, electrons correlations should play a secondary role, and not be able to localise holes into a Vervey-Mott insulator. Using a simplified model where the Coulomb intraatomic energy U is neglected but for magnetic properties where its effects are treated in Hartree-Fock perturbation, we describe the electronic structures of long range commensurate AF. We point out that, for large enough doping this should be replaced by some sort of short range AF, with a characteristic pseudogap. We think that this is observed in hole doped compounds such as YBCO with commensurate AF, where the peak in $T_c$ occurs when the Fermi level crosses the lower energy peak of this pseudogap. This explains the large value of observed $T_c$, but also the symmetry of the superconductive gap (d for underdoped, d + ip for overdoped samples, where p varies proportionnal to overdoping). In this approach, the symmetry of the gap is related to the special symmetry of the wave functions in an near the pseudogap and coherent with an isotropic attractive phonon mediate interaction, while an interaction mediated by AF fluctuations would not lead to any appreciable gap.

It is expected that the same scheeme could apply to electron doped compounds, where O2p shells should also have holes in appreciable amounts ; superconductivity should develop in a range of short range AF, with its characteristic features of a pseudogap. The superconductive gap should again have a basic $d_{x^2-y^2}$ symmetry, with an ip character increasing proportional to overdoping. Here too, the optimum $T_c$ should take place when the (here AB) peak of the pseudogap crosses the Fermi level.

If these qualitative predictions are confirmed, the theoretical picture should be completed. More quantitative estimates of    and of the magnetic moments could give a more precise estimate of $U_e$. Electron correlations could then be explicitely included [22], and the coupling between CuO2 planes should be included [2] to actually estimate $T_c$. From that point of view, it is rather surprising that the recent experiments on doping of apparently a single $CuO_2$ plane by electric voltage could give a large and well defined drop of resistivity at $T_c$, contrary to Kosterlitz and Thouless's predictions.

Two major difficulties for quantitative estimates relate to the short range AF order. First, what is exactly its self consistent fine scale structure and the corresponding variations of the density of states in the pseudogap for the AF commensurate with the lattice considered here [32] ? What is the current description of superconductivity in LSCO ? Finally, from the point of view developed here, it is difficult anyway to analyse in detail the electronic structure in the normal state above $T_c$. It is true that, in hole underdoped samples, the peak B of the pseudogap is expected to be related to a feature somewhat broadened by magnetic disorder but very flat in energy in the reciprocal space : this seems to be observed in the excitation of holes [18] [23]. A less marked feature is expected in overdoped hole samples, corresponding to electron excitations to the same peak of the pseudogap and a (fainter) anomaly is expected in the scheeme along the line $T_p$, fig. 3. Similar observations should be possible in the electron doped samples. The analogy with long range AF certainly suggests electronic Hall conduction for holes overdoping and hole Hall conduction for electrons overdoping, in agreement with observations [5]. But the switch of sign of Hall conduction when going from over the underdoping remains to be explained in detail. Also the exact form of the Fermi surface and its broadening by magnetic disorder remain to be worked out, especially for underdoped samples.

It can be noted also that a long range AF is often observed near to a superconductive one [24]. For our specific mechanism to work, a fringe of short range AF should be observed next to the long range one, in the doping or pressure conditions where superconductivity is observed.

We can compare in fine our approach with other somewhat similar ones, also using a delocalised electrons picture for the cuprates.

- Our approach, leading to a dominant superconductive d gap, is in practice very near to Pines' approach [15], although the detail of the local densities associated with the pseudogap should be introduced in his computations. Also, as we start from a phonon coupling, we could account for a (small) isotope effect. Finally the small is component predicted for overdoping is not present in his picture.

- Our own previous approaches [2], [4] assumed AF short range, leading then, in our simplified picture, to a phonon coupling and superconductivity with an s gap due to the Van Hove anomaly pictured in punctuated line, fig. 3. This should have been dominant on the d coupling by AF fluctuations, leading to a first order transition of the gap near optimal doping in YBCO [21]. This is not observed, and the presence of AF short range order in the overdoped range observed by neutrons [2] as well as by NMR techniques in pure and doped samples of YBCO [28] contradicts our initial assumption.

- The band structure of delocalised Cu3d and O2p electrons renormalised for large repulsions U [20] contains two Van Hove anomalies when    is large and a (smaller) transfer integral is added between second neighbours Cu. This reproduces a density of states analogous to that of the pseudogap g , fig. 3 ; and the changes with electrons or holes dopings of the Hall conductivity and of the thermal variation of the electrical resistivity can be satisfactorily explained in that way [31]. A similar analysis could certainly be made in our model, with similar conclusions for the thermal variation of the electrical resistivity ; and we have pointed out that, at least in the overdoped regime, our model also predicts the sign of the Hall conduction correctly. There is an obvious analogy between the two models. Our own takes into account the AF short range order and the pseudogap, that reference [30] does not consider. Our model can also explain in a natural way the low temperature increase of resistivity in the underdoped regime, as the electronic states involved in the pseudogap are at least partly localised.

- Mrkonjc and Barisic [11] have deduced renormalised LCAO parameters from the Fermi surface of YBCO as deduced from photoemission spectra. With the use of , t and a direct transfer t between oxygens, a good agreement is obtained for very small values of  and t (smaller than t ). The authors argue that, in the covalent regime defined in our text, renormalisation due to U leads indeed to smaller values of  and t, as we also have pointed out for t [2]. However the very small values deduced for  and t could only be obtained for practically infinite values of U, as in the tJ model [22]. This seems somewhat extreme when compared with values in transitional metals : in cuprates as in metals, $U_e$, when deduced from the AF properties or from X rays spectra, seems less than the width of the electronic band and of order t. As the Fermi surface is analysed without reference to short range AF order, more discussion is probably in order. It is clear than equ. (6) which should approximately replace equ. (1) in our model strongly contracts the energy scale near the AF gap.

In conclusion, there are definite similarities in a number of delocalised scheemes proposed more or less recently ; and it is only the general fit of predicted properties with observations that can allow to choose between them. We are aware that our own model, being so far very qualitative, fullfills difficultly this condition.

*The authors are especially grateful for stimulating discussions on delocalised electrons scheemes with D. Pines, S. Barisic, J. Bok, J. Bouvier and L. Gorkov, together for experimental results with G. Deutscher, H. Alloul, Y. Petrov, S.W. Loram, P. Bourges, C.C. Tsuei, M. Lagües and B. Batlog. G. Deutscher has spotted an initial error in the symmetry of the imaginary gap. Thanks are also due to C. Fradin and N. Dupuis for typing the text and drawing the figures.*

# TITLE OF FIGURES

1.- Qualitative symmetry of AF and superconductive (S) phases versus doping z in cuprates.

2.- Fermi surface for non magnetic compounds. 0 zero doping (square ABCD) ; e electron doping ; h hole doping.

3.- Antiferromagnetic gap g and pseudogap g̃. B bonding and AB antibonding peaks in the density of states. Dotted curve : van Hove singularity of non magnetic state. Abscissae : energy E and doping z. $E < E_o$ hole doping ; $E > E_o$ electron doping.

4.- Stability versus doping z of commensurate ($Q_o$) and uncommensurate ($\tilde{Q}\ Q_o$) AF long range order. Dotted curve : stability of commensurate short range AF (the position of $z_o$ and $\tilde{z}_o$ for short range commensurate order $\tilde{Q}_o$ marked on the abscissae).

5.- Schematic phase diagram for hole doped cuprates. AF and S : long range antiferromagnetic and supraconductive phases (as in fig. 1) ; $z_o$ optimum doping ; $T_P$ (and possibly $\tilde{T}_P$) : temperatures of characteristic excitations from (and to) the pseudogap binding peak associated with the short range AF.

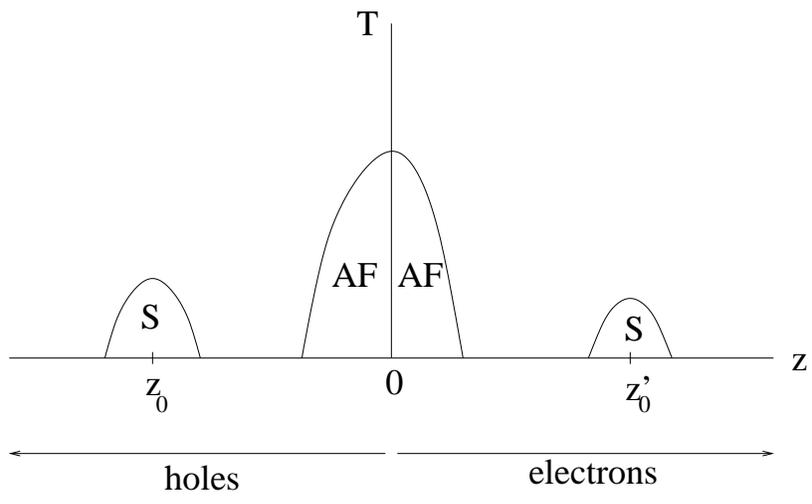

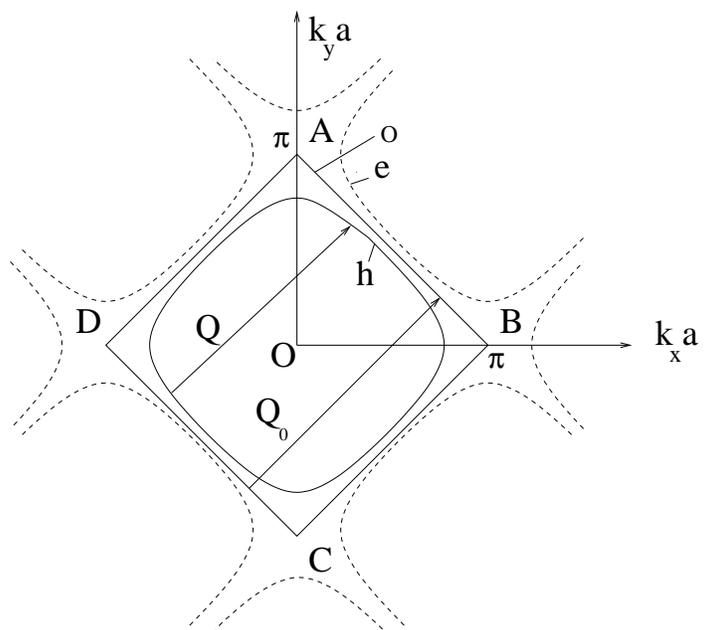

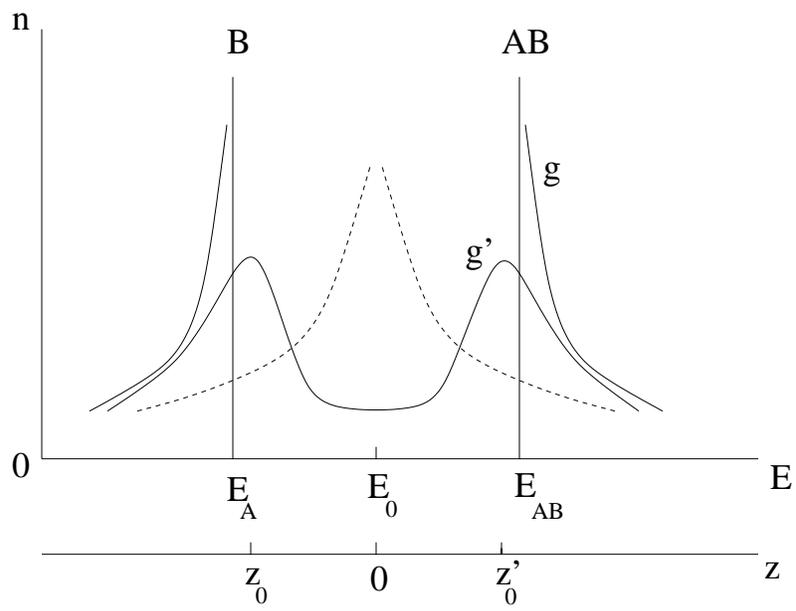

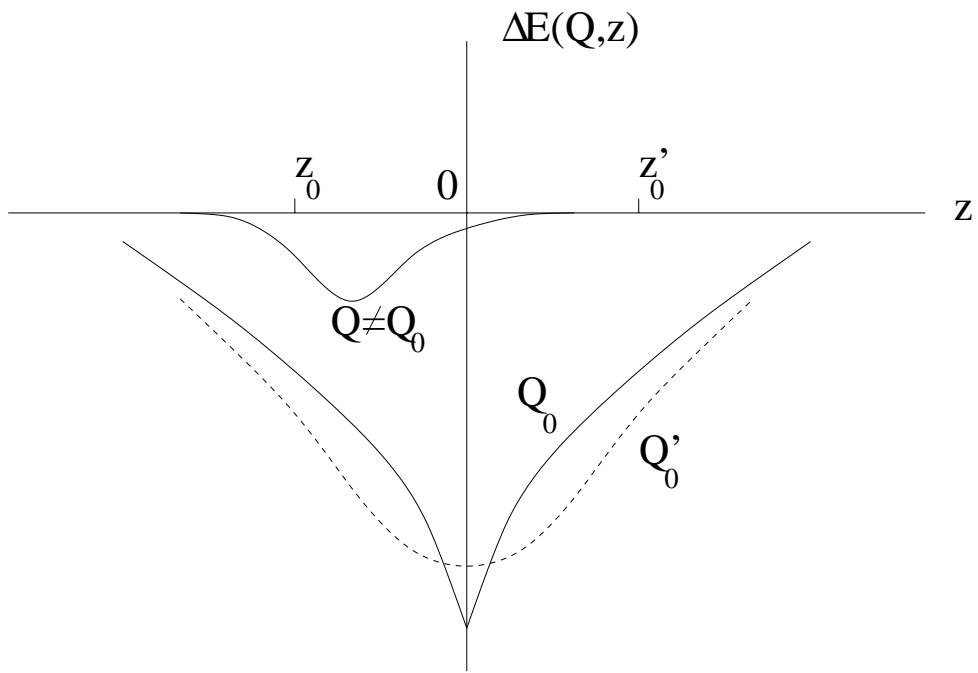

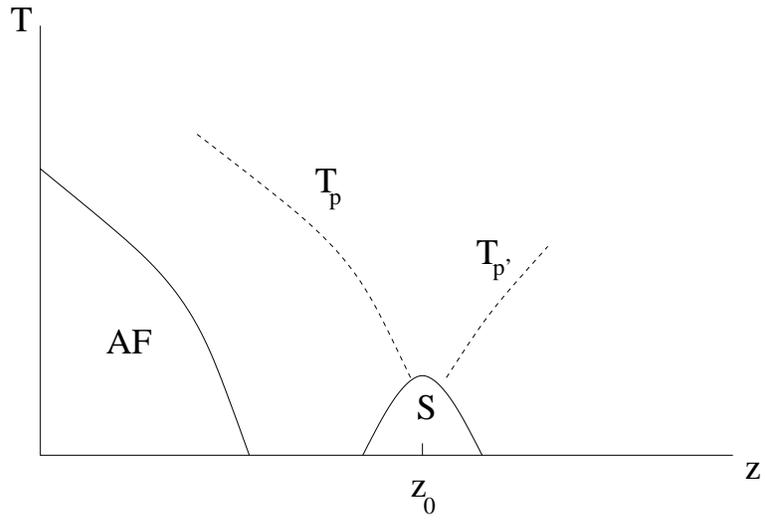